\title{ \textbf{Dynamic Analysis of a Predator and Prey Model with Some Computational Simulations}}
\author{ Sarbaz H. A.  Khoshnaw  
\\ Department of Mathematics, University of Raparin,\\ Kurdistan Region of Iraq}  
\documentclass[a4paper,12pt]{article}      
	\usepackage{multirow,booktabs,colortbl,tabularx}
\usepackage{makeidx}
\usepackage{latexsym}         
\usepackage{mathrsfs}                 
\usepackage{url}                                              
 \usepackage{enumitem}                                                                 
 \usepackage[titletoc]{appendix}                                                     
\usepackage{tikz}                                                  
\usetikzlibrary{shapes,arrows,positioning} 
\usepackage{cite}  
\usepackage{relsize}  
              
\usepackage{setspace}    
\usepackage{amsmath, amssymb, amsthm}
\usepackage{chemarr, graphicx, subfigure} 
\usepackage{amsmath, amsthm, amssymb, amsfonts}
\usepackage{multirow,booktabs,colortbl,tabularx}    
\usepackage{resizegather}         
 \usepackage{cite} 
 \usepackage{makeidx}        
\usepackage{latexsym}             
\usepackage{mathrsfs}                    
\usepackage{url}
  \usepackage{enumitem}
  \usepackage{tikz}                                                  
\usetikzlibrary{shapes,arrows,positioning}
  \tikzstyle{decision} = [diamond, draw, fill=blue!20, text width=3cm, text badly centered, node distance=3cm, inner sep=0pt]
\tikzstyle{block} = [rectangle, draw, fill=blue!20,text width=18em, text centered, rounded corners, minimum height=3em]
\tikzstyle{line} = [draw, -latex]     
 \medskip       
 
\begin{document}                                                   
 \maketitle        
\begin{abstract}
Mathematical modelling and numerical simulations of interaction populations are crucial topics in systems biology. The interactions of ecological models may occur among individuals of the same species or individuals of different species. Describing the dynamics of such models occasionally requires some techniques of model analysis. Choosing appropriate techniques of model analysis is often a difficult task. We define a prey (mouse) and predator (cat) model. The system is modelled by a pair of non-linear ordinary differential equations using mass action law, under constant rates. A proper scaling is suggested to minimize the number of parameters. More interestingly, we propose a homotopy technique with $n$ expanding parameters for finding some analytical approximate solutions. Furthermore, using the local sensitivity
method is another important step forward in this study because it helps to identify critical model
parameters. Numerical simulations are provided using Matlab for different parameters and initial conditions.
\end{abstract}

\maketitle  
\section{Introduction}
The main problem of models for interacting populations was observed by Umberto D'Ancona. He performed a statistical analysis of the fish that were sold in the markets of Trieste, Fiume, and Venice between 1910 and 1923. Umberto showed that this coincided with increases in the relative frequency of some species and decreases in the relative frequency of other species \cite{Kot2001}. 

Lotka in 1925 proposed a mathematical model for the population dynamics of a predator and prey from a hypothetical chemical reaction. One year after, Volterra independently suggested a simple model for the predation of one species by another to explain the oscillatory levels of certain fish catches in the Adriatic.
 The proposed equations have become a well known model of mathematical biology. The model is composed of a pair of differential equations that describe predator and prey dynamics in their simplest case \cite{lotka1925}. Then, the model was further developed to include density dependent prey growth and a functional response of the form developed by C.S. Holling. The developed model has become known as the Rosenzweig--McArthur model. Both the Lotka--Volterra and Rosenzweig--MacArthur models have been used to explain the dynamics of natural populations of predators and prey. There are some examples of natural populations of predators and prey such as the lynx and snowshoe hare data of the Hudson Bay Company and the moose and wolf populations in Isle Royale National Park \cite{ Gilpin1973, Goel1971, Jost2005}.\\  

The model was further studied and developed in terms of stability analysis \cite{Murray2002}. He considered the generalised Lotka--Volterra system where there are $k$ prey species and $k$ predators. He also classified three main situations of population interactions. The first situation is that the populations are in a predator--prey type when the growth rate of one population is decreased and the other increased. We have also a competition situation. This occurred when the growth rate of each population is decreased. Then, the third situation is called mutualism or symbiosis, if each population's growth rate is enhanced.

Recently, a variety of non-linear differential equations has been solved by perturbation methods. The homotopy perturbation method (HPM) is a series expansion method. This is used to calculate analytical approximate solutions for non-linear ODEs. The method is based on an assumption that a small parameter must exist in non-linear equations \cite{He2000,He1999,He2003,Chowdhury,Vogt}. A homotopy perturbation method with two expanding parameters was suggested by \cite{He2014}.\\
  
 Representing biological processes formally for mathematical modelling are crucial topics in systems biology. They are given by using the classical theory of chemical kinetics \cite{brigs1925, semenoff1939, segel1989}. There are some studies of chemical kinetics which are involved in determining biochemical processes in systems biology, for example, dynamic and static limitation in multiscale reaction networks \cite{gorbanredul2008}; a symptotology of chemical reaction networks \cite{gorbanraduzin2010}; robust simplifications of multiscale biochemical networks \cite{radulescugorzi2008}; reduction of dynamical biochemical reaction networks in computational biology \cite{radulescugorzi2012}; a model reduction method for biochemical reaction networks \cite{rao2014}; iterative approximate solutions of kinetic equations for reversible enzyme reactions \cite{khoshnaw2013}; reduction of a kinetic model of active export of importins \cite{khoshnaw2015a}; model reductions in biochemical reaction networks \cite{khoshnaw2015b} and identifying critical parameters in SIR model for
spread of disease \cite{khoshnaw2017}.\\  
 
We assume that an ecological model consists of       
\begin{description}[noitemsep,nolistsep]          
\item[$\bullet$] A vector of species   
 $\mathcal{S}=(\mathcal{S}_{1} , \mathcal{S}_{2}, ... , \mathcal{S}_{m})$,   
 for each species $\mathcal{S}_{j}$, $j=1,2,...,m$ a non negative variable $x_{j}$ is defined. In other words, $x_{j}$ is the population of species $\mathcal{S}_{j}$.   
\item[$\bullet$] A vector of population interactions ( ecological reaction rates) 
 $\mathcal{V}=(v_{1}, v_{2}, ... , v_{n}).$    
\item[$\bullet$] A vector of constant interactions between two species   
$\mathcal{K}=(k_{1}^{\mp}, k_{2}^{\mp}, ... , k_{n}^{\mp})$.         
\end{description}
\noindent Stoichiometric equations for $n$ elementary reversible reactions are given below:  
\begin{equation} 
\begin{array}{llll}  
 \mathlarger{\mathlarger{\sum\limits}}_{j=1}^m \alpha_{ij} \mathcal{S}_{j} \underset{k_{i}^{-}}{ \overset{k_{i}^{+}}{\rightleftharpoons}}\mathlarger{\mathlarger{\sum\limits}}_{j=1}^m \beta_{ij} \mathcal{S}_{j}, \quad i=1,2,...,n .
\end{array}
\label{elementry}  
 \end{equation}  
The non-negative integers $\alpha_{ij}$ and $\beta_{ij}$ are called stoichiometric coefficients. The standard mass action law is used to define the rate of reactions. The reaction rates are:
\begin{equation}
\begin{array}{llll}  
 v_{i} = k_{i}^{+} \mathlarger{\mathlarger{\prod\limits}}_{j=1}^m x_{j}^{\alpha_{ij}}(t)-k_{i}^{-} \mathlarger{\mathlarger{\prod\limits}}_{j=1}^m x_{j}^{\beta_{ij}}(t), \quad i=1,2,...,n ,
\end{array}
\label{rates}
 \end{equation}
where $k_{i}^{+} > 0$ and $k_{i}^{-} \geq 0$ are the reaction rate coefficients. \\         
The stoichiometric matrix is $\mathcal{G}=(\gamma_{ij}),$ where $\gamma_{ij}=\beta_{ij}-\alpha_{ij},$ for $ i=1,2,...,n$ and   $j=1,2,..,m$. The stoichiometric vector $\gamma_{i}$ is the \textit{i}th row of $\mathcal{G}$ with coordinates $\gamma_{ij}=\beta_{ij}-\alpha_{ij}$.
\noindent The system of ODE describes the dynamics of chemical reactions. The kinetic equations are:
\begin{equation}
\begin{array}{llll}  
\dfrac{d\mathcal{X}}{dt}=\mathcal{W}(\mathcal{X}(t), \mathcal{K})=\mathcal{G} \enskip \mathcal{V}(\mathcal{X}(t), \mathcal{K}), \\
\mathcal{X}(0)=\mathcal{X}_{0}, \quad  t \in I\subset \mathbb{R}^{+}\cup \lbrace 0\rbrace, 
\end{array}
\label{eq1} 
 \end{equation}  
\noindent where $\mathcal{G}$ is a stoichiometric matrix of $m$ by $n$, $\mathcal{X}(0)$ is a vector of initial populations. The kinetic equations (\ref{eq1}) can also be expressed as follows:
\begin{equation}
\begin{array}{llll}  
\dfrac{d\mathcal{X}}{dt}=\mathlarger{\mathlarger{\sum\limits}}_{i}\gamma_{i}v_{i} .
\end{array}
\label{eq1a} 
 \end{equation}  
 
 The main contribution in this work is to apply some mathematical tools to simplify and analyse the mouse and cat model and then identify the model elements (variables and parameters). We propose a number of steps of model analysis, which plays a role in reducing the number of elements and in calculating analytical approximate solutions of the model. The proposed steps and their advantages are simply given. The first step is that we use mass action law to define the model by a pair of non--linear ordinary differential equations, under constant rates. Then, a proper scaling is used in order to minimize the number of elements. This becomes a good step forward for simplifying the original model. Another step is calculating some analytical approximate solutions of the simplified model using a homotopy technique. Furthermore, we simulate the model populations for different values of the remaining parameter $\mu$ in two and three dimensional planes. Generally, we can conclude that for different value of $\mu$ there is a different dynamic of the model. Interestingly, the population of predators (cats) becomes more stable when the value of $\mu$ becomes larger. Finally, we use the local sensitivity
method in this study. This helps us to identify critical model parameters of the reduced model.

\section{Mouse and Cat Model} 
\begin{figure}[ht]         
      \begin{center}    
        \subfigure{%
            \label{fig:third}      
            \includegraphics[width=0.9\textwidth]{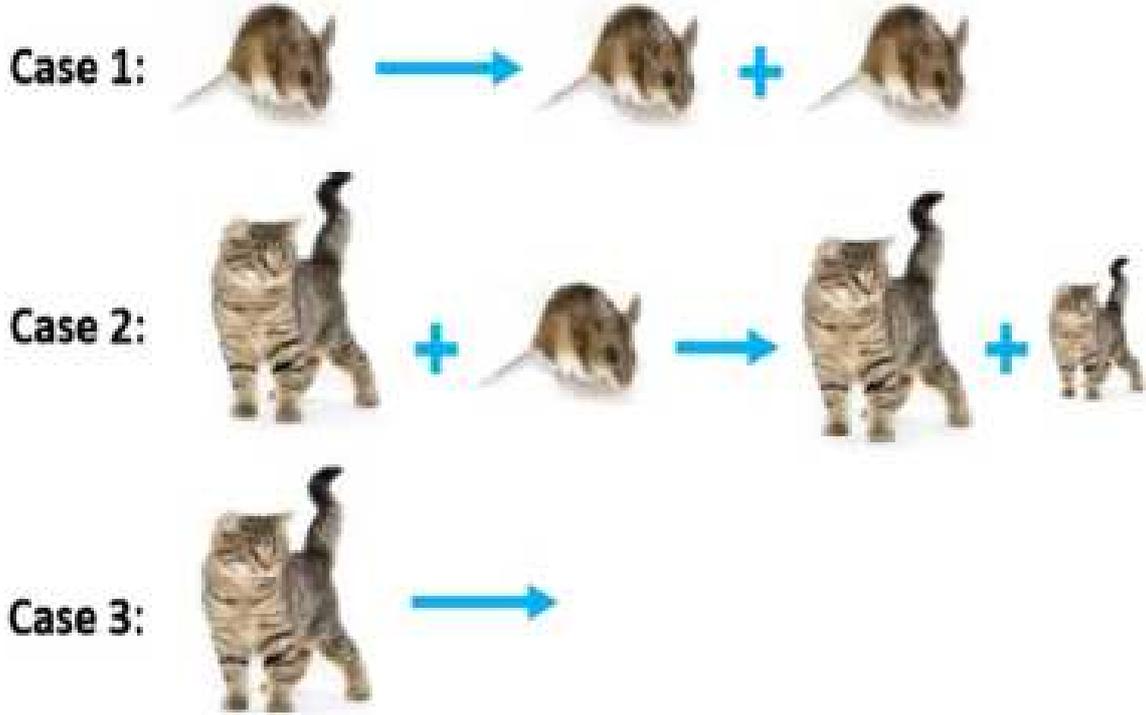}
        }
    \end{center}     
\caption {Three mechanisms of a simple mouse and cat model.}  
   \label{f1}  
\end{figure}

 Consider the predator and prey model, there are two species, one as a prey (mouse) and the other as a predator (cat). The model can simply be given by three mechanisms of population interactions; see Figure \ref{f1}.   

The simplest model can be shown as a set of chemical reactions as bellow
\begin{center} 
\begin{equation} 
\begin{array}{llll}  
M {\overset{\alpha}\longrightarrow } 2M, \\
M+C {\overset{\beta}\longrightarrow } (1+\delta)C, \\ 
C{\overset{\gamma}\longrightarrow }...
\end{array}
\label{eqy} 
 \end{equation}     
 \end{center} 
\noindent where $M$ denotes the population of the prey (mice) species, and $C$ denotes
the population of the predator (cats) species. $\alpha, \beta, \gamma$ and $\delta$ are positive real parameters describing the interaction of two species. The given set of positive model parameters are defined in Table \ref{table1}.
\begin{table}   
  \center
  \begin{tabular}{ | l  | l | l | l |  }
    \hline
     \textbf {Parameters} & \textbf {Definitions}   \\ \hline      
     $\alpha$ &  The growth rate of mouse    \\ \hline
     $\beta$ &  The rate at which cats destroy mice     \\ \hline
      $\gamma$& The death rate of cats    \\ \hline
    $\delta$&  The rate at which cats increase by consuming mice     \\ \hline       
    \end{tabular}
    \caption{The model parameters and their definitions.}
    \label{table1}   
\end{table}   
\section{Mathematical Formulation}
We use the idea of classical chemical kinetics in order to define a mathematical model for the set of chemical mechanisms (\ref{eqy}). Therefore, a set of stoichiometric vectors of the system can be given as  
\begin{gather*}
  \setlength{\arraycolsep}{1\arraycolsep}
  \text{$
  \gamma_{1}  =\begin{pmatrix}
 1\\
 0
\end{pmatrix},
\quad  \gamma_{2} =\begin{pmatrix}
-1\\
 \delta
\end{pmatrix},
\quad \gamma_{3} =\begin{pmatrix}
 0\\
-1
\end{pmatrix}$}.
\end{gather*} 
We use mass action law (\ref{rates}) to define the chemical reaction rates of the model\\
$v_{1}=\alpha M ,\quad  v_{2}=\beta MC,\quad  v_{3}=\gamma C .$\\

\noindent The kinetic equations are given as
\begin{equation}
\begin{array}{llll}  
\dfrac{d}{dt}\text{$
\begin{pmatrix}
 M\\
 C
\end{pmatrix}$}=\mathlarger{\mathlarger{\sum\limits}}_{i=1}^{3}\gamma_{i}v_{i}.
\end{array}
\label{eqk1} 
 \end{equation}
Then the system of differential equations takes the form 
\begin{equation}
\begin{array}{llll}  
\dfrac{d M}{dt}=\alpha M - \beta M C,\\
\\
\dfrac{d C}{dt}=\delta \beta M C - \gamma C,
\end{array}
\label{eqk2} 
 \end{equation}
with initial populations $M(0)=M_{0}$ and $C(0)=C_{0}$. \\

\noindent In mathematical modelling, scaling of variables is an essential task for model reduction. There would be more than one way to scale the variables. Particularly, differently scaled equations and differently reduced models can be obtained by different variable and parameter scales. We can scale the variables to minimize the number of parameters. A simple scaling for the system (\ref{eqk2}) is used. This is by introducing the following new variables 
\begin{gather*}
M^{*}=\dfrac{\delta}{\gamma}M, \quad C^{*}=\dfrac{\beta}{\alpha}C \quad \text{and}\quad t^{*}=\sqrt{\alpha \gamma}t .  
\end{gather*}
Thus, the system of differential equations (\ref{eqk2}) becomes
 \begin{equation}
\begin{array}{llll}  
\dfrac{d M^{*}}{dt^{*}}=\mu \big(M^{*} -M^{*} C^{*}  \big),\quad M^{*}(0)=M^{*}_{0},\\
\\
\dfrac{d C^{*}}{dt^{*}}=\dfrac{1}{\mu} \big(M^{*} C^{*} - C^{*} \big),\quad  C^{*}(0)=C^{*}_{0},
\end{array}
\label{eqk3} 
 \end{equation}
where $\mu=\sqrt{\dfrac{\alpha}{\gamma}}.$  \\
From the system (\ref{eqk3}), we can calculate an implicit analytical solution for the model  
species. This is given below:
 \begin{equation}
\begin{array}{llll}    
M^{*} e^{-M^{*}}=C^{*}{^{-\mu^{2}}}e^{\mu^{2} C^{*} -\phi},
\end{array}
\label{eqk4}   
 \end{equation}
where  $\phi=M^{*}_{0}-\mu^{2}C^{*}_{0} -\ln(M^{*}_{0} C^{*}_{0}{^{\mu^{2}}})$. 
\section{ A Homotopy Technique with \textit{n} Expanding Parameters}
In this study, we introduce a homotopy perturbation technique with \textit{n} expanding parameters. To explain the basic ideas of the technique, we consider the following non-linear ODEs with $n$-1 non-linear terms:   
\begin{equation}  
\begin{array}{llll}  
L(u)+\sum\limits_{i=1}^{n-1}N_{i}(u)=0,
\end{array}\label{NHTP1}
\end{equation}
\noindent where $u\in R^{m}$, $m\geqslant 1$, $L$ is a linear operator, $N_{i}$ is a non-linear operator for $i=1,2,...,n$-1. We construct the following homotopy equation:
\begin{equation}  
\begin{array}{llll}
\tilde{L}(u)+\sum\limits_{i=1}^{n-1}q_{i}N_{i}(u)+q_{n}\big[L(u)-\tilde{L}(u)\big]=0,
\end{array}\label{NHTP2}
\end{equation}  
\noindent where $q_{i}$ is homotopy parameter, $q_{i}\in[0,1]$ for $i=1,2,...,n$, $\tilde{L}$ is a linear operator and $\tilde{L}(u)=0$ can approximately describe the main property of equation \ref{NHTP1}.\\
\noindent The approximate solutions of equation \ref{NHTP2} can be expressed as a power series in $q_{1},q_{2},...,q_{n}$
\begin{equation}  
\begin{array}{llll}
u=u_{0}+\sum\limits_{i=1}^{n}q_{i}u_{i}+\sum\limits_{i=1}^{n}\sum\limits_{j=1}^{n}q_{i}q_{j}u_{ij}+\sum\limits_{i=1}^{n}\sum\limits_{j=1}^{n}\sum\limits_{k=1}^{n}q_{i}q_{j}q_{k}u_{ijk}+...
\end{array}\label{NHTP3}
\end{equation}  
\noindent Setting $q_{i}=1$ for $i=1,2,...,n$, we obtain the solution as follows:
\begin{equation}  
\begin{array}{llll}
u=u_{0}+\sum\limits_{i=1}^{n}u_{i}+\sum\limits_{i=1}^{n}\sum\limits_{j=1}^{n}u_{ij}+\sum\limits_{i=1}^{n}\sum\limits_{j=1}^{n}\sum\limits_{k=1}^{n}u_{ijk}+...
\end{array}\label{NHTP4}
\end{equation}
\noindent The proposed technique can be illustrated by the simplified model (\ref{eqk3}).  A homotopy of the system (\ref{eqk3}) can be constructed with given expanding parameters:
\begin{equation}
\begin{array}{llll}  
\dfrac{d M^{*}}{dt^{*}}-\mu M^{*}+\mu q_{1} M^{*} C^{*}=0  ,\quad M^{*}(0)=M^{*}_{0},\\
\\
\dfrac{d C^{*}}{dt^{*}}+\dfrac{1}{\mu} C^{*}- \dfrac{1}{\mu} q_{1} M^{*} C^{*}=0  ,\quad  C^{*}(0)=C^{*}_{0},
\end{array}
\label{eqk5} 
 \end{equation}
 
 \noindent The approximate solution of system (\ref{eqk5}) is introduced in the forms
\begin{equation}  
\begin{array}{llll}
M^{*}=M^{*}_{0}+\sum\limits_{i=1}^{2}q_{i}M^{*}_{i}+\sum\limits_{i=1}^{2}\sum\limits_{j=1}^{2}q_{i}q_{j}M^{*}_{ij}+\sum\limits_{i=1}^{2}\sum\limits_{j=1}^{2}\sum\limits_{k=1}^{2}q_{i}q_{j}q_{k}M^{*}_{ijk}+...\\
C^{*}=C^{*}_{0}+\sum\limits_{i=1}^{2}q_{i}C^{*}_{i}+\sum\limits_{i=1}^{2}\sum\limits_{j=1}^{2}q_{i}q_{j}C^{*}_{ij}+\sum\limits_{i=1}^{2}\sum\limits_{j=1}^{2}\sum\limits_{k=1}^{2}q_{i}q_{j}q_{k}C^{*}_{ijk}+...
\end{array}\label{eqk6}
\end{equation} 

\noindent The above equations can be easily written:
\begin{equation}  
\begin{array}{llll}
M^{*}=M^{*}_{0}+q_{1}M^{*}_{1}+q_{2}M^{*}_{2}+q_{1}^{2}M^{*}_{3}+q_{1}q_{2}M^{*}_{4}+q_{2}^{2}M^{*}_{5}+...
 \\ 
C^{*}=C^{*}_{0}+q_{1}C^{*}_{1}+q_{2}C^{*}_{2}+q_{1}^{2}C^{*}_{3}+q_{1}q_{2}C^{*}_{4}+q_{2}^{2}C^{*}_{5}+...
\end{array}\label{eqk7} 
\end{equation}

\noindent Substituting equations (\ref{eqk7}) into equations (\ref{eqk5}) and collecting the same power of $q_{1}^{n_{1}}q_{2}^{n_{2}}$ $(n_{1},n_{2}=0,1,2,...,n)$ and setting coefficients to zero, the following systems are obtained:
\begin{equation}  
\begin{array}{llll}
q_{1}^{0}q_{2}^{0}:\quad \dfrac{dM^{*}_{0}}{dt^{*}}- \mu M^{*}_{0}=0, \quad M^{*}_{0}(0)=M^{*}_{0} \\
q_{1}^{1}q_{2}^{0}:\quad \dfrac{dM^{*}_{1}}{dt^{*}}-\mu M^{*}_{1}+\mu M^{*}_{0} C^{*}_{0}=0, \quad M^{*}_{1}(0)=0 \\
q_{1}^{0}q_{2}^{1}:\quad \dfrac{dM^{*}_{2}}{dt^{*}}-\mu M^{*}_{2}=0, \quad M^{*}_{2}(0)=0 \\
q_{1}^{2}q_{2}^{0}:\quad \dfrac{dM^{*}_{3}}{dt^{*}}-\mu M^{*}_{3}+\mu M^{*}_{0} C^{*}_{1}+\mu M^{*}_{1} C^{*}_{0}=0, \quad M^{*}_{3}(0)=0 \\
q_{1}^{1}q_{2}^{1}:\quad \dfrac{dM^{*}_{4}}{dt^{*}}-\mu M^{*}_{4}+\mu M^{*}_{0} C^{*}_{2}+\mu M^{*}_{2} C^{*}_{0}=0, \quad M^{*}_{4}(0)=0 \\
q_{1}^{0}q_{2}^{2}:\quad \dfrac{dM^{*}_{5}}{dt^{*}}-\mu M^{*}_{5}=0, \quad M^{*}_{5}(0)=0 \\
\end{array}\label{eqk8} 
\end{equation}
and
\begin{equation}  
\begin{array}{llll}
q_{1}^{0}q_{2}^{0}:\quad \dfrac{dC^{*}_{0}}{dt^{*}}- \dfrac{1}{\mu} C^{*}_{0}=0, \quad C^{*}_{0}(0)=C^{*}_{0} \\
q_{1}^{1}q_{2}^{0}:\quad \dfrac{dC^{*}_{1}}{dt^{*}}+\dfrac{1}{\mu} C^{*}_{1}-\dfrac{1}{\mu} M^{*}_{0} C^{*}_{0}=0, \quad C^{*}_{1}(0)=0 \\
q_{1}^{0}q_{2}^{1}:\quad  \dfrac{dC^{*}_{2}}{dt^{*}}+\dfrac{1}{\mu} C^{*}_{2}=0, \quad C^{*}_{2}(0)=0\\
q_{1}^{2}q_{2}^{0}:\quad  \dfrac{dC^{*}_{3}}{dt^{*}}+\dfrac{1}{\mu} C^{*}_{3}-\dfrac{1}{\mu} M^{*}_{0} C^{*}_{1}-\dfrac{1}{\mu} M^{*}_{1} C^{*}_{0}=0, \quad C^{*}_{3}(0)=0 \\
q_{1}^{1}q_{2}^{1}:\quad \dfrac{dC^{*}_{4}}{dt^{*}}+\dfrac{1}{\mu} C^{*}_{4}-\dfrac{1}{\mu} M^{*}_{0} C^{*}_{2}-\dfrac{1}{\mu} M^{*}_{2} C^{*}_{0}=0, \quad C^{*}_{4}(0)=0\\
q_{1}^{0}q_{2}^{2}:\quad \dfrac{dC^{*}_{5}}{dt^{*}}+\dfrac{1}{\mu} C^{*}_{5}=0, \quad C^{*}_{5}(0)=0\\
\end{array}\label{eqk9}     
\end{equation}
Systems (\ref{eqk8}) and (\ref{eqk9}) can be analytically solved for $M^{*}_{i}(t^{*})$ and $C^{*}_{i}(t^{*})$ respectively where $i=0,1,2,3,4,5$. Then, the analytical approximate solution of the model (\ref{eqk3}) is obtained using the suggested technique
 \begin{equation}  
\begin{array}{llll}
M^{*}(t^{*})=\lim\limits_{\substack{q_{1} \to 1 \\ q_{2} \to 1}}M^{*}(t^{*})= \sum\limits_{i=0}^{5}M^{*}_{i}(t^{*}),\\
C^{*}(t^{*})=\lim\limits_{\substack{q_{1} \to 1 \\ q_{2} \to 1}}C^{*}(t^{*})= \sum\limits_{i=0}^{5}C^{*}_{i}(t^{*}).
\end{array}\label{eqk10}
\end{equation}
\section{Numerical Simulations}
\begin{figure}[ht]         
      \begin{center}    
        \subfigure{%
            \label{fig:third}     
            \includegraphics[width=1\textwidth]{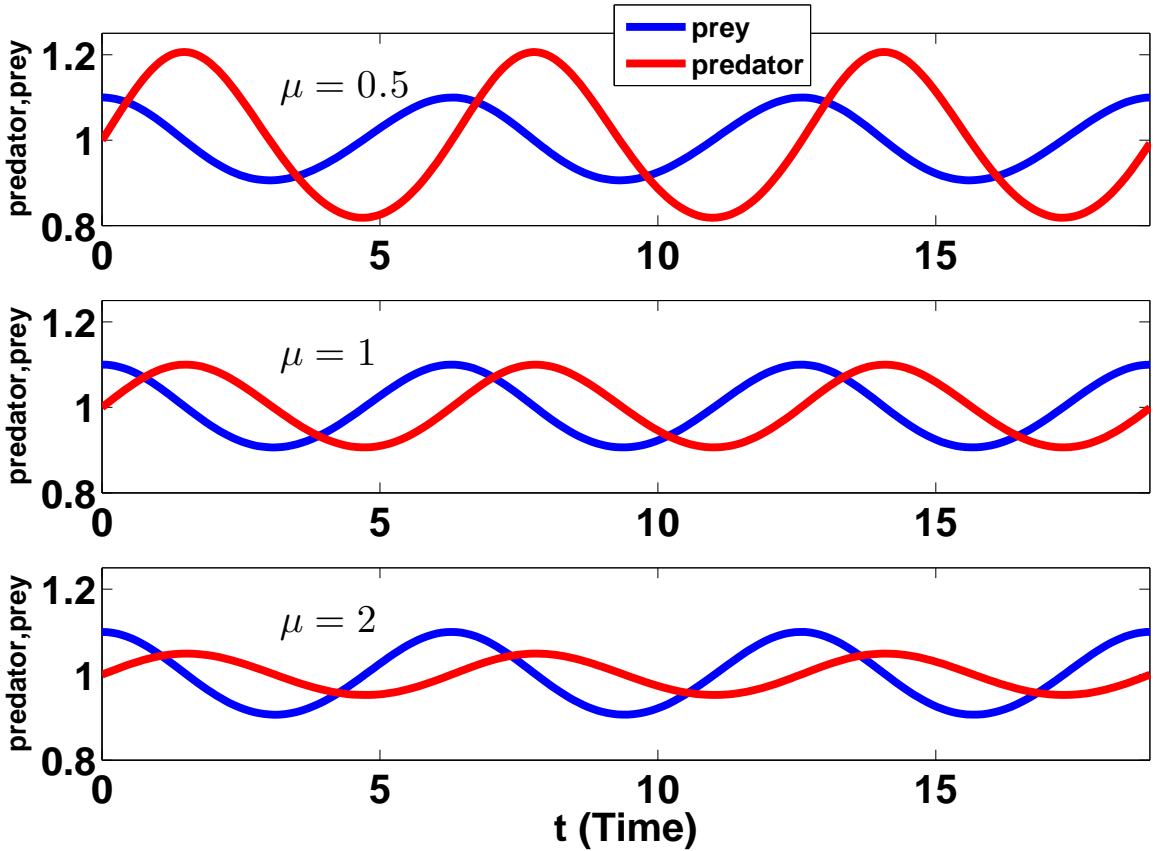}
        }
    \end{center}       
\caption {Numerical simulations for the simplified model (\ref{eqk3}) using Matlab for the remaining parameter $\mu = 0.5 ,1, 2$  with initial conditions $M^{*}(0)=1.1$ and $C^{*}(0)=1$ .}  
   \label{f2}  
\end{figure}

\begin{figure}[ht]         
      \begin{center}    
        \subfigure{%
            \label{fig:third}        
            \includegraphics[width=1\textwidth]{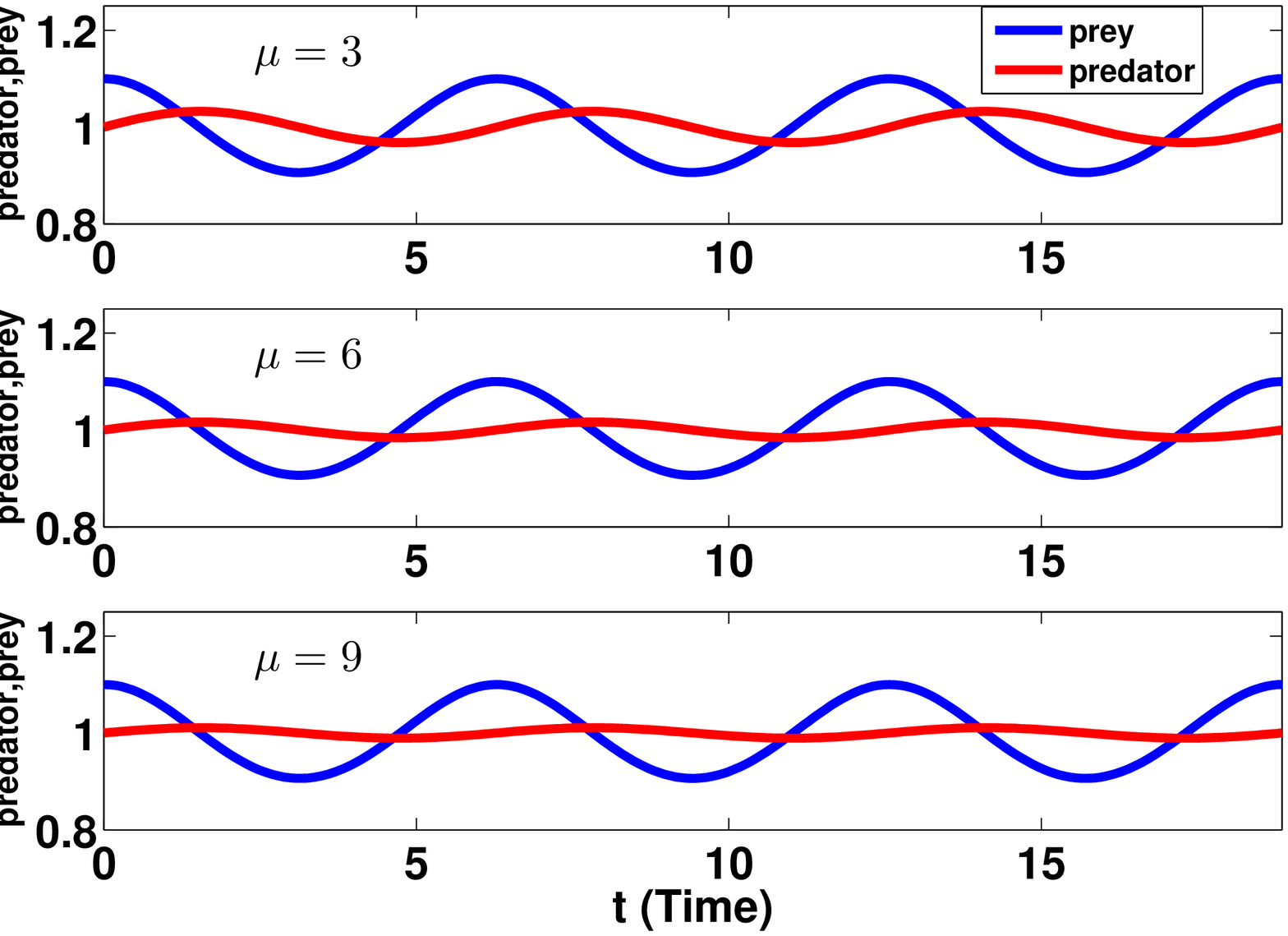}
        }
    \end{center}       
\caption {Numerical simulations for the simplified model (\ref{eqk3}) using Matlab for the remaining parameter $\mu = 3 ,6, 9$  with initial conditions $M^{*}(0)=1.1$ and $C^{*}(0)=1$ .}  
   \label{f3}  
\end{figure}

\begin{figure}[ht]         
      \begin{center}    
        \subfigure{%
            \label{fig:third}     
            \includegraphics[width=0.9\textwidth]{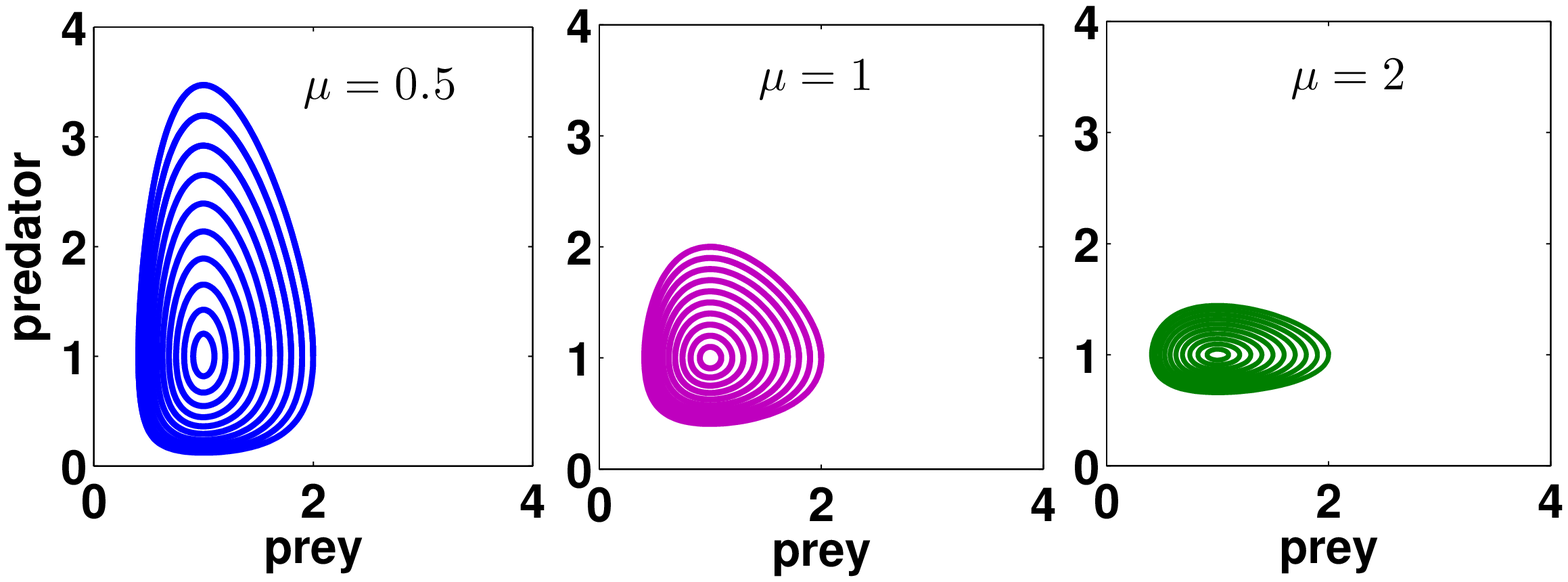}
        }
    \end{center}       
\caption {Numerical simulations for the simplified model (\ref{eqk3}) using Matlab for different values of the remaining parameter $\mu = 0.5 ,1, 2$  with initial conditions $M^{*}(0)=1.1$ and $C^{*}(0)=1$ .}  
   \label{f4}  
\end{figure}

\begin{figure}[ht]          
      \begin{center}    
        \subfigure{%
            \label{fig:third}        
            \includegraphics[width=0.9\textwidth]{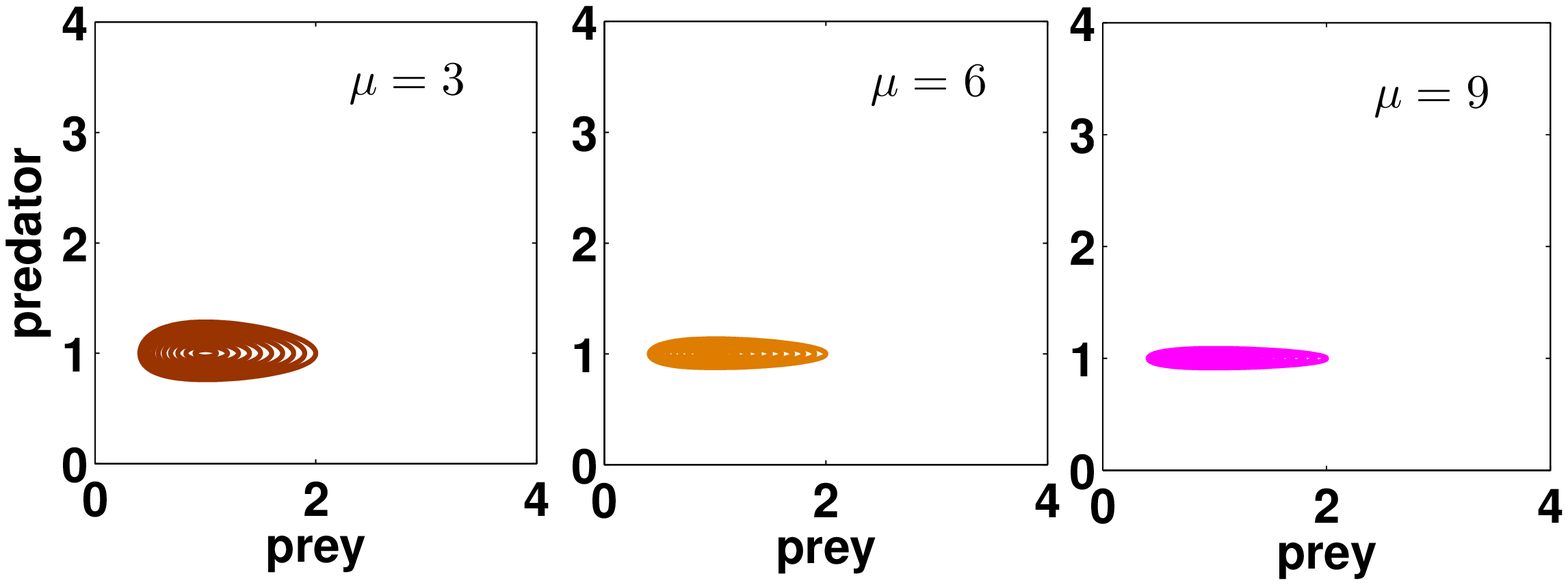}
        }
    \end{center}       
\caption {Numerical simulations for the simplified model (\ref{eqk3}) using Matlab for different values of the remaining parameter $\mu = 3 ,6, 9$  with initial conditions $M^{*}(0)=1.1$ and $C^{*}(0)=1$ .}  
   \label{f5}  
\end{figure}

\begin{figure}[ht]         
      \begin{center}    
        \subfigure{%
            \label{fig:third}        
            \includegraphics[width=0.6\textwidth]{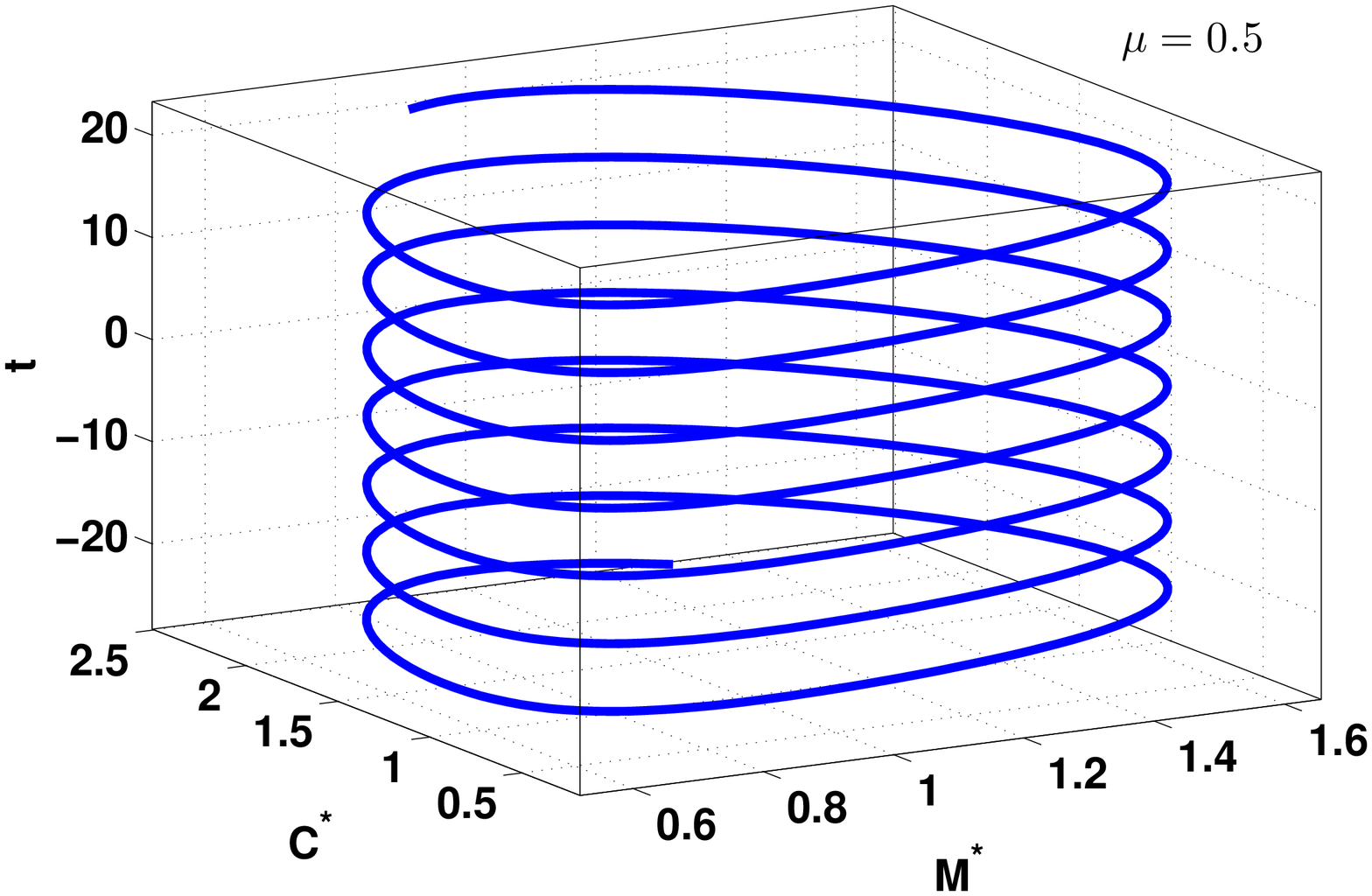}
        }\\    
        \subfigure{%
            \label{fig:third}        
            \includegraphics[width=0.6\textwidth]{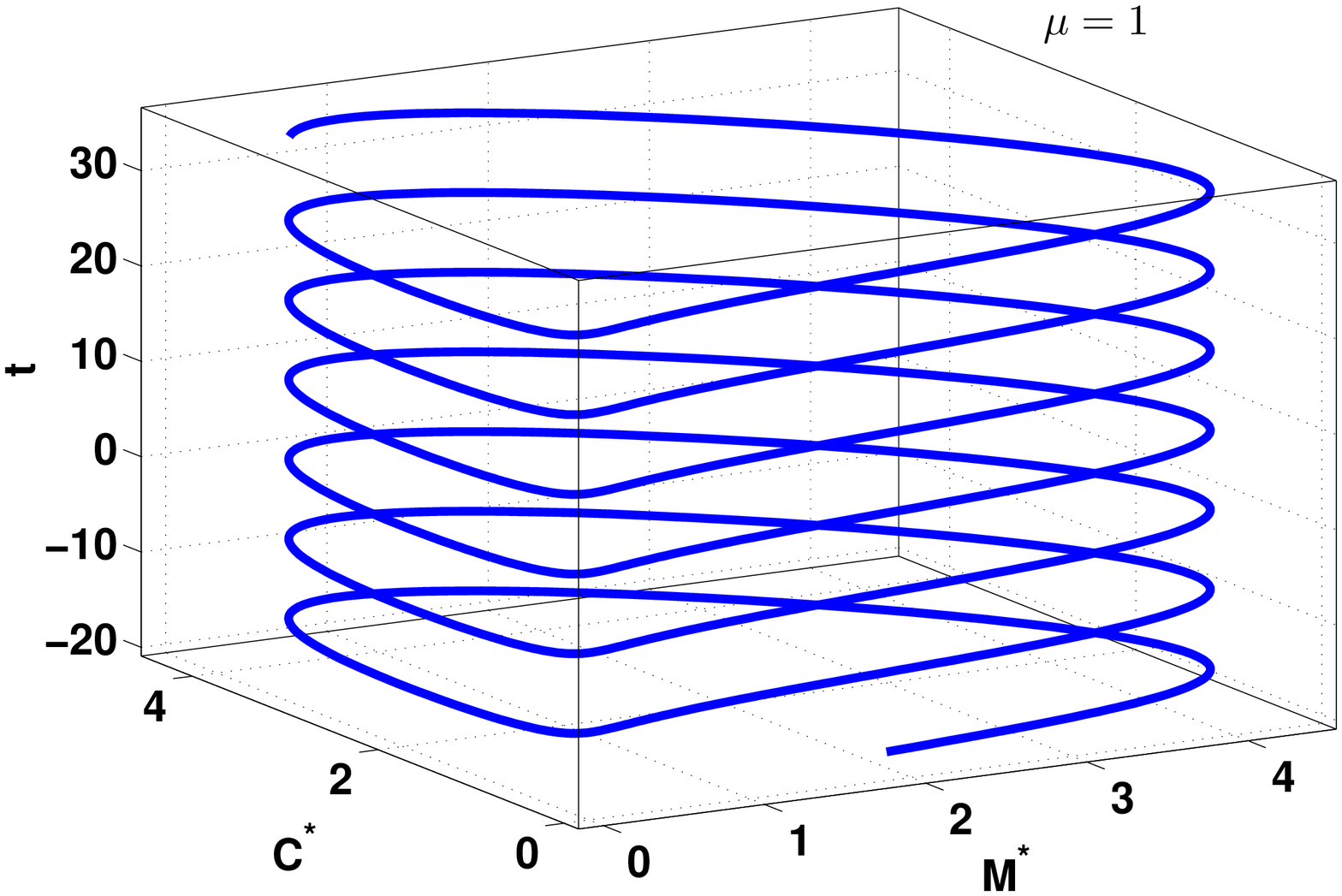}
        }\\    
        \subfigure{%
            \label{fig:third}        
            \includegraphics[width=0.6\textwidth]{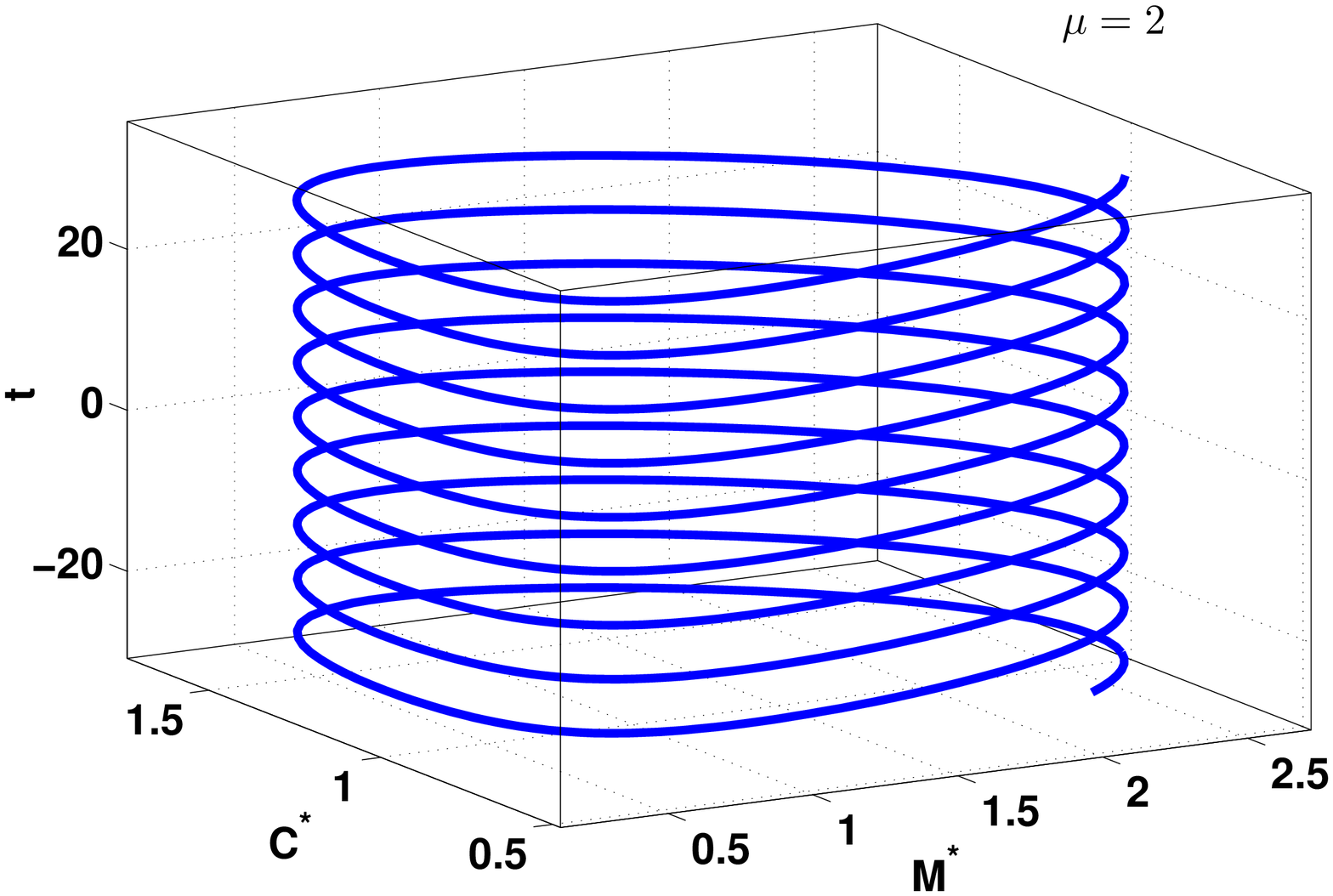}
        }
    \end{center}         
\caption {Dynamics of the simplified model (\ref{eqk3}) and numerical simulations using  Pplane8 for Matlab in three dimensional plane, for different values of parameter $\mu$ with initial conditions $M^{*}(0)=1.1$ and $C^{*}(0)=1$.}  
   \label{f8}    
\end{figure}

 The numerical approximate solutions of the simplified model (\ref{eqk3}) for different values of $\mu$ can be expressed in Figures \ref{f2}--\ref{f8}. The numerical simulations are computed in Matlab. We simulate the model populations for different values of the remaining parameter $\mu$ in two and three dimensional planes. Generally, we can conclude that for different value of $\mu$ there is a different dynamic of the model. Interestingly, the population of predators becomes more stable when the value of $\mu$ becomes larger.
 
\clearpage

\clearpage
\clearpage
\section{Sensitivity Analysis}
\noindent The idea of sensitivity analysis has been used in dynamic analysis of biochemical kinetics and ecological  models. This method is used to determine which variable or parameter is sensitive to a particular condition which is defined by a variable or parameter. The system of ODEs discussed here is: 
\begin{equation}  
\begin{array}{llll}
\dfrac{dy_{i}}{dt}=\phi_{i}\big(Y(t), \mathcal{K} \big), \quad i=1, 2, ..., m.
\end{array}\label{SA}  
\end{equation}
\noindent The model input $\mathcal{K}$ is a vector of parameters, and the model output $Y$ is a vector of state variables. Local sensitivity is the changes in state variables $y_{i}$, $i=1,2,..,m$ with respect to parameters $k_{p}$, $p=1,2,...,n$ . \\ 
             
\noindent The general form of the local sensitivity is given as a Jacobian matrix as follows
\begin{equation}  
\begin{array}{llll}
\dot{\mathcal{S}}=\phi_{k_{p}}+\mathcal{J} . \mathcal{S}, \quad p=1,2,...,n,
\end{array}\label{SA5}
\end{equation}
\noindent where the matrices $\mathcal{S}, \phi_{k_{p}}$ and $\mathcal{J}$ are defined by  
\begin{gather*}
  \setlength{\arraycolsep}{1\arraycolsep}
  \text{$
 \mathcal{S}
=\begin{bmatrix}
\frac{\partial y_{1}}{\partial k_{p}}\\
\frac{\partial y_{2}}{\partial k_{p}}\\
\vdots \\
 \frac{\partial y_{m}}{\partial k_{p}}\\
\end{bmatrix}, 
 \quad \phi_{k_{p}}  =\begin{bmatrix}
\frac{\partial \phi_{1}}{\partial k_{p}} \\
\frac{\partial \phi_{2}}{\partial k_{p}} \\
\vdots \\
\frac{\partial \phi_{m}}{\partial k_{p}} 
\end{bmatrix}
, \quad \mathcal{J}=  \begin{bmatrix}
   \frac{\partial \phi_{1}}{\partial y_{1}} & \frac{\partial \phi_{1}}{\partial y_{2}} & \cdots &  \frac{\partial \phi_{1}}{\partial y_{m}}   \\  
  \frac{\partial \phi_{2}}{\partial y_{1}} & \frac{\partial \phi_{2}}{\partial y_{2}} & \cdots &  \frac{\partial \phi_{2}}{\partial y_{m}}   \\ 
  \vdots & \vdots & \ddots & \vdots \\
   \frac{\partial \phi_{m}}{\partial y_{1}} & \frac{\partial \phi_{m}}{\partial y_{2}} & \cdots &  \frac{\partial \phi_{m}}{\partial y_{m}}  \\      
    \end{bmatrix} .   
$}        
\end{gather*}
\noindent The initial conditions of the equation (\ref{SA5}) are determined by the input parameter $k_{p}$ and the initial condition of the output variables $y_{i}$.\\

\begin{figure}[ht]          
      \begin{center}    
        \subfigure[$\mu = 50$]{%
            \label{fig:third}        
            \includegraphics[width=0.43\textwidth]{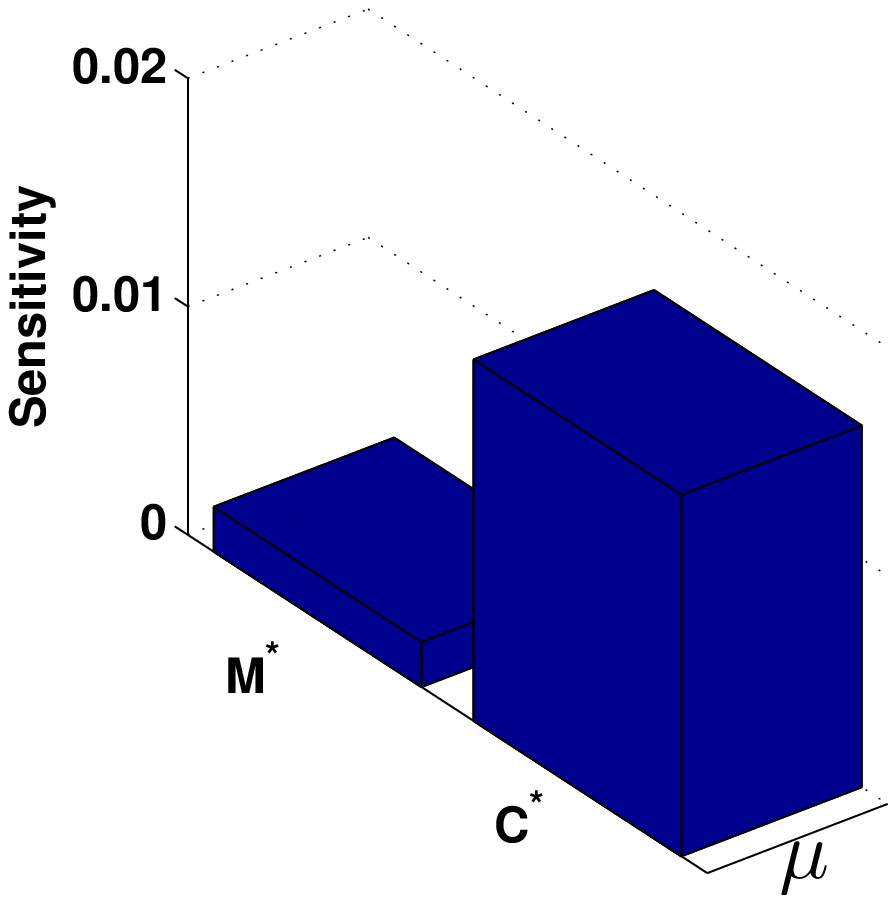}
        }
        \subfigure [$\mu = 150$]{%
            \label{fig:third}        
            \includegraphics[width=0.43\textwidth]{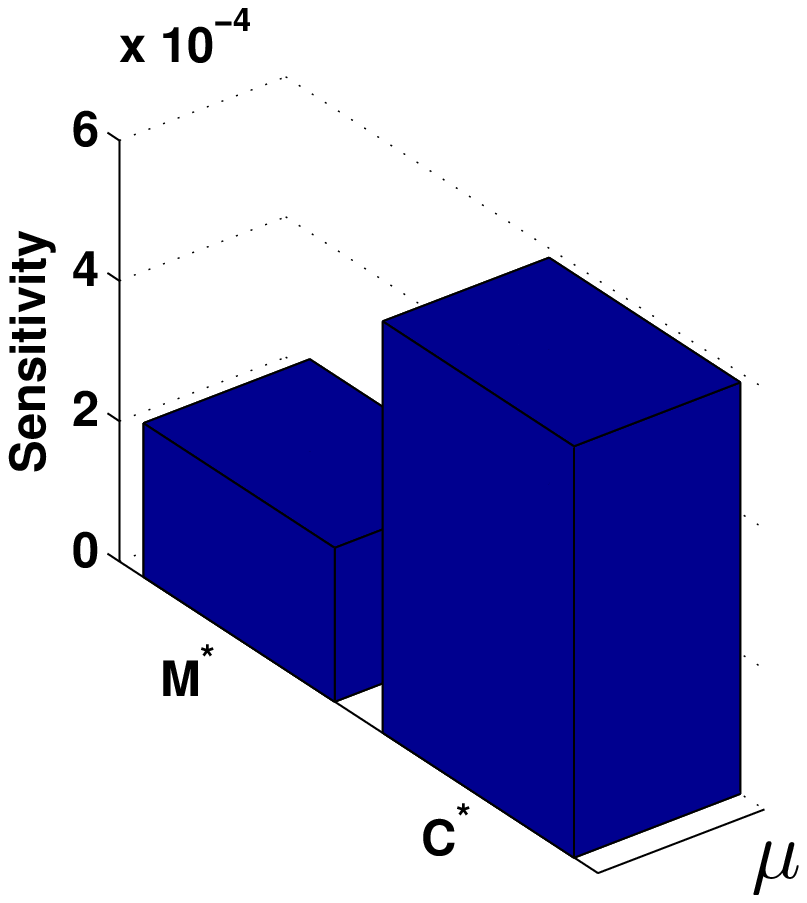}
        }\\
        \subfigure [$\mu = 400$]{%
            \label{fig:third}        
            \includegraphics[width=0.43\textwidth]{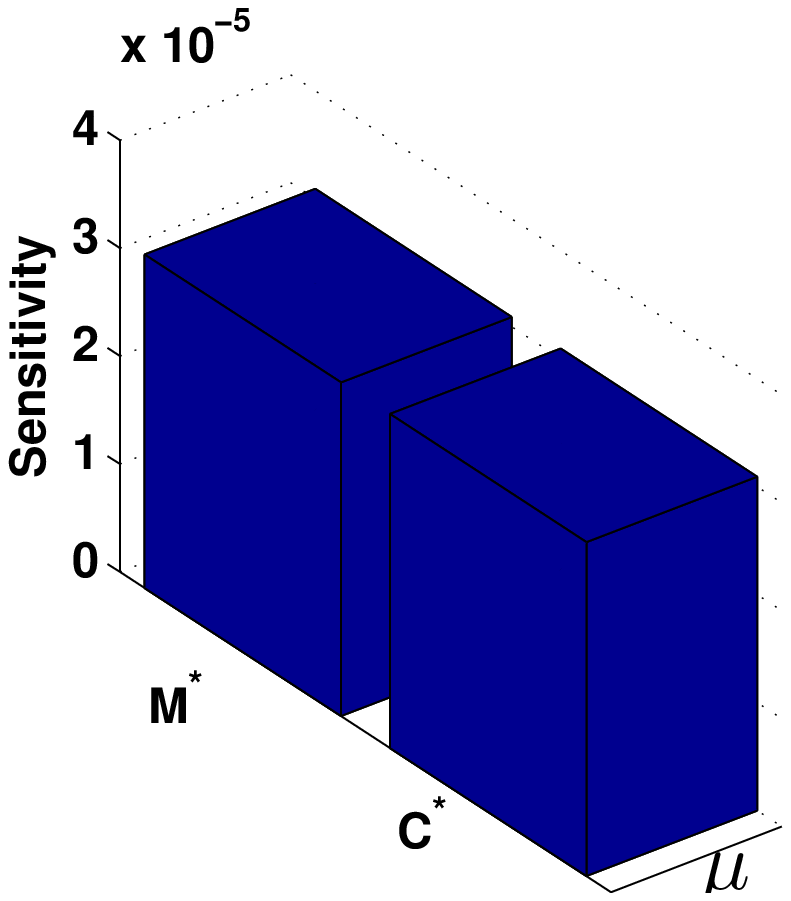}
        }
        \subfigure [$\mu = 500$]{%
            \label{fig:third}        
            \includegraphics[width=0.43\textwidth]{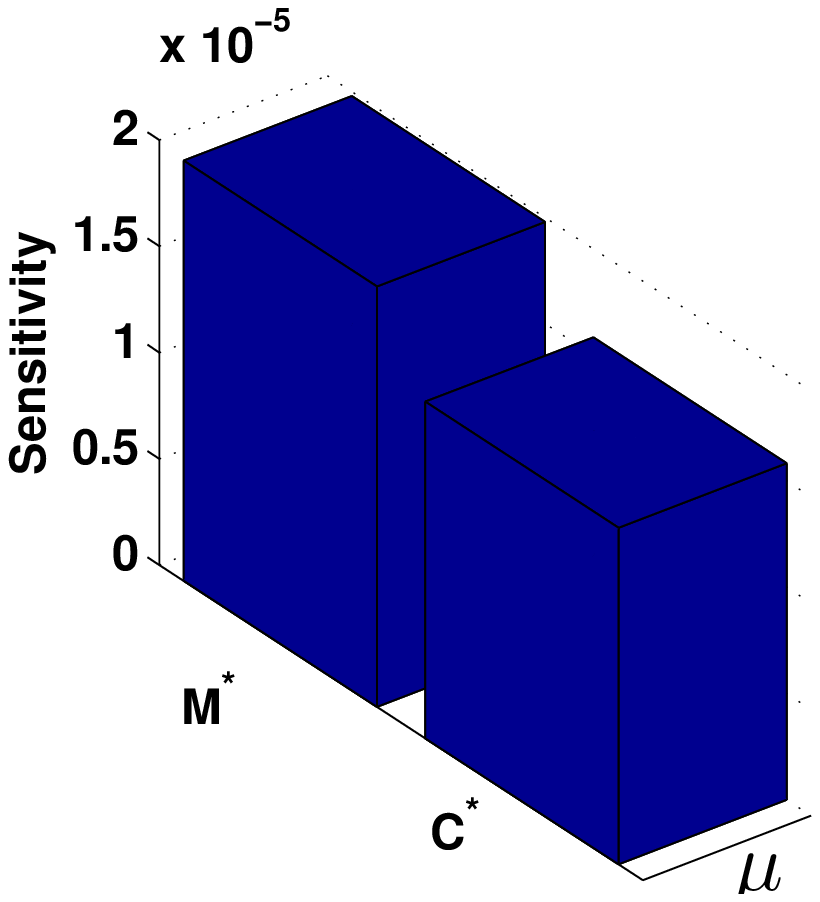}
        }\\
         \subfigure [$\mu = 550$]{%
            \label{fig:third}        
            \includegraphics[width=0.43\textwidth]{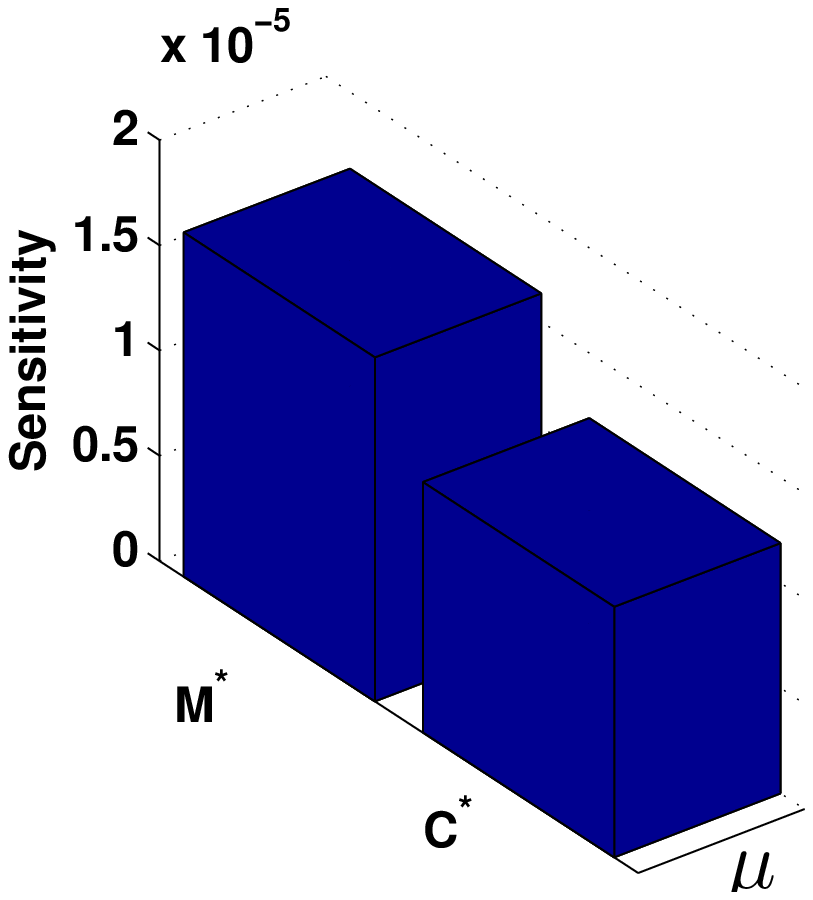}
        }
        \subfigure[$\mu = 3000$] {%
            \label{fig:third}        
            \includegraphics[width=0.43\textwidth]{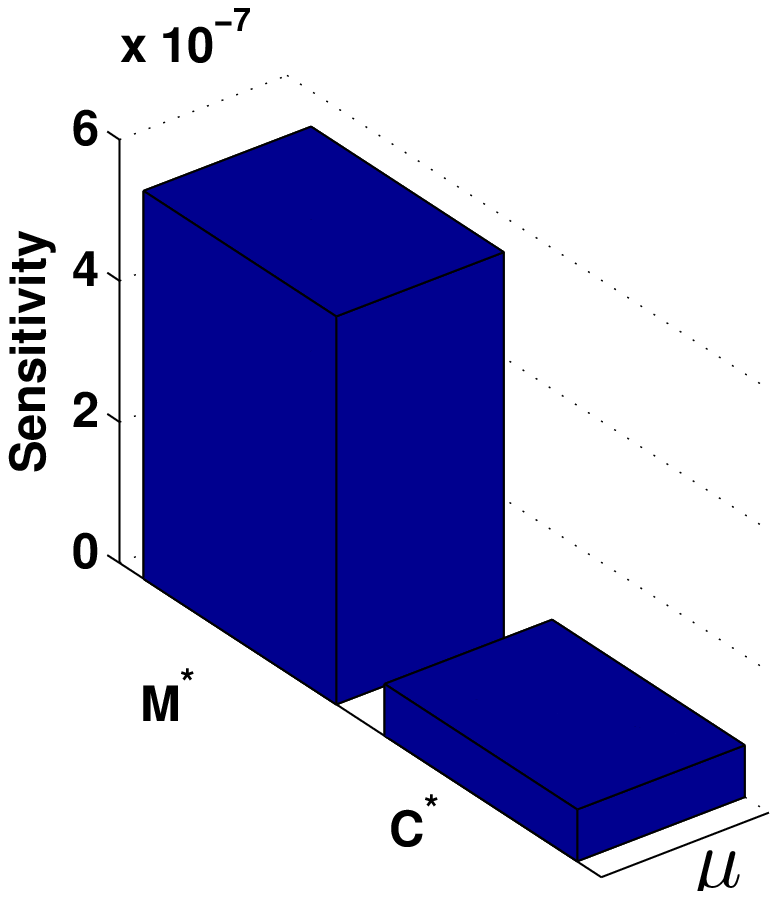}
        }
    \end{center}         
\caption {The local sensitivity of state variables $M^{*}$ and $C^{*}$ with respect to the given parameter
$\mu$ for different values, using the SimBiology Toolbox for Matlab in the time interval [0,10]
units of time with initial populations $M^{*}(0)=100$ and $C^{*}(0)=20$.}  
   \label{S1}    
\end{figure}

We calculate the local sensitivity of state variables $M^{*}$ and $C^{*}$ of the system (\ref{eqk3}) with respect to the given parameter $\mu$ to identify critical mode parameters. We identify that the population of predators (cats) is more sensitive to the remaining parameter $\mu$ when $\mu \in (0 , 400)$ while it is less sensitive to the given parameter $\mu$ as $\mu > 400$. More interestingly, both populations (predators and preys) have the same sensitivity to $\mu$ when $\mu=400$ ; see Figure \ref{S1}.
Results here are computed in numerical simulations using the SimBiology Toolbox for Matlab in the time interval [0,10] units of time. Identifying critical model parameters in this study is a good step forward for describing and understanding the model dynamics of interacting populations.  
\clearpage

\clearpage
\section{Conclusions} 
We have studied a prey and predator model with two species and four parameters. The system is modelled using mass action law and classical chemical kinetics under constant rates. A proper scaling is used in this study in order to reduce the number of parameters. Variable scaling here importantly plays in minimizing the number of model parameters from four parameters to only one parameter. We apply the local sensitivity method to identify critical model parameters. A homotopy perturbation technique with $n$ expanding parameters is proposed. The method gives some analytical approximate solutions of the simplified model. This becomes a good step forward in different ways. Firstly, the simplified model is a system of non linear ordinary differential equations that can not be solved exactly. Secondly, the approximate solutions help us to understand global dynamics of the model. Furthermore, the proposed method can further be developed and applied to high dimensional non linear ecological models. \\
 
\noindent Simulations and results in this study are obtained using Matlab for different values of the remaining parameter $\mu$ and initial populations. Results show some interesting points. The first point is that for different value of $\mu$ there is a different dynamic of the model in two and three dimensional planes. Another point is that the population of predators (cats) becomes more stable when the value of $\mu$ becomes larger. More interestingly, the population of predators (cats) is more sensitive to the remaining parameter $\mu$ when $\mu < 400$ while it is less sensitive to the parameter $\mu$ as $\mu > 400$. Furthermore, both populations (predators and preys) have the same sensitivity to $\mu$ when $\mu=400$.
Finally,the results in this paper could be accurate, robust, and easily applied by ecologists for various purposes, such as reproducing ecological data and identifying critical ecological model parameters. The proposed techniques here will be applied to a wide range of complex ecological interaction populations.


\end{document}